\documentclass[twocolumn,aps,prc,showpacs,superscriptaddress]{revtex4}
\usepackage{epsfig}
\newcommand {\bw}	{\begin{widetext}}
\newcommand {\ew}	{\end{widetext}}
\newcommand {\be}	{\begin{equation}}
\newcommand {\ee}	{\end{equation}}
\newcommand {\bea}	{\begin{eqnarray}}
\newcommand {\eea}	{\end{eqnarray}}
\newcommand {\mn}[1]	{\langle#1\rangle}
\newcommand {\mne}[1]	{\langle e^{in(#1)}\rangle}

\newcommand {\M}[1]	{\mn{#1}}
\newcommand {\mM}[1]	{\mn{\M{#1}}}
\newcommand {\vv}[1]	{v\{{\rm #1}\}}
\newcommand {\vtwo}[1]	{v_2\{{\rm #1}\}}

\newcommand {\ncl}[1]	{N_{\rm cl#1}}

\newcommand {\Fi}	{\phi_i}
\newcommand {\Fj}	{\phi_j}
\newcommand {\Fk}	{\phi_k}
\newcommand {\Fl}	{\phi_{\ell}}
\newcommand {\Fm}	{\phi_{\mu}}
\newcommand {\Fn}	{\phi_{\nu}}
\newcommand {\Fp}	{\phi'_i}

\newenvironment{color}[3]
{
}{
}

\begin{document}

\title{Possibility to disentangle anisotropic flow, flow fluctuation, and nonflow assuming Gaussian fluctuations}
\author{Li Yi}
\affiliation{Department of Physics, Purdue University, West Lafayette, Indiana 47907, USA}
\author{Fuqiang Wang}
\affiliation{Department of Physics, Purdue University, West Lafayette, Indiana 47907, USA}
\author{Aihong Tang}
\affiliation{Brookhaven National Laboratory, Upton, New York 11973, USA}

\begin{abstract}
We suggest the possibility to disentangle anisotropic flow, flow fluctuation, and nonflow using two-, four-, and six-particle azimuthal moments assuming Gaussian fluctuations. We show that such disentanglement is possible when the flow fluctuations are large, comparable to the average flow magnitude. When fluctuations are small, the disentanglement becomes difficult. We verify our results with a toy-model Monte Carlo simulation.
\end{abstract}

\pacs{25.75.-q, 25.75.Dw}

\keywords{heavy-ion; flow; flow fluctuation; nonflow}

\maketitle

\section{Introduction}

Azimuthal distributions of hadrons produced in non-central heavy-ion collisions are anisotropic~\cite{Ollitrault_ecc,flowreview,STAR_flow,PHENIX_flow}. The distributions are often expressed by Fourier series,
\be dN/d\phi\propto 1+\sum_{n=1}^{\infty}2v_n\cos(n\phi)\,,\label{eq:fourier}\ee
where $\phi$ is particle azimuthal angle relative to the reaction plane. 
Coefficients $v_n$ quantify the azimuthal anisotropies of different harmonics. 
%
Since reaction plane is not experimentally accessible, experiments use particle azimuthal correlations 
to measure anisotropies~\cite{Wang_2part,v2method}. 
As a result, the measured anisotropies using two-particle correlation method, $\vv{2}$, contain not only 
flow but also flow fluctuations and nonflow~\cite{Borghini_nonflow,Miller_fluc,STAR_nonflow,Voloshin_fluc,Ollitrault_fluc_nonflow}. 

Flow fluctuations are mainly due to initial geometry fluctuations~\cite{Miller_fluc}. 
Flow fluctuations could also result from fluctuating responses~\cite{Stephanov} of the collision system to an identical initial condition. Flow fluctuations cause a variation of anisotropy from event to event in a sample of supposedly the same events. We note, however, experimentally one cannot ensure the selection of the same events a priori, so flow fluctuations come also partially from real differences in the collisions, e.g.~of different impact parameters~\cite{Miller_fluc}.

Nonflow is caused by particle correlations that are not directly related to the reaction plane, such as jet-correlations and resonance decays~\cite{Borghini_nonflow}. 
In general, nonflow is a few-body correlation while 
flow is a many-body correlation. It was suggested that nonflow is largely canceled in a particular combination of two- and four-particle correlations~\cite{Borghini_multipart,Borghini_multipart2}. The resultant anisotropic flow from this four-particle method, $\vv{4}$, contains therefore significantly less nonflow than that from the two-particle method, $\vv{2}$~\cite{STAR_v4,STAR_v2}. 

Elliptic flow, $v_2$, arising from the elliptical overlap region of the colliding nuclei, has been extensively studied~\cite{STAR_whitepaper,PHENIX_whitepaper,PHOBOS_whitepaper}. Significant efforts have been invested to study $v_2$ fluctuations and nonflow~\cite{GangWang,Alver_nonGaus,PHOBOS_nonflow,Sorensen1,Sorensen2}. Triangular flow, $v_3$, can arise from the triangularity of the initial overlap geometry~\cite{Alver,Zulum,Petersen,Qin,Schenke,Teaney,Ko}, and may be dominated by fluctuations. 
So far flow fluctuations and nonflow have not been experimentally distinguished~\cite{Sorensen1,Sorensen2}.
It is well known that the two- and four-particle azimuthal moments are not enough to determine the average flow, flow fluctuation, and nonflow~\cite{Voloshin_fluc}. In this article we argue that it may be possible to determine the three quantities by the additional information of six-particle azimuthal moment. We support our argument by a toy-model Monte Carlo simulation.

\section{Azimuthal Moments}

The two-, four-, and six-particle $n^{\rm th}$ azimuthal moments are~\cite{Borghini_multipart}:
\bea
\M{2}&=&\mne{\Fi-\Fj}\,,\label{eq:m2}\\
\M{4}&=&\mne{\Fi+\Fj-\Fk-\Fl}\,,\label{eq:m4}\\
\M{6}&=&\mne{\Fi+\Fj+\Fk-\Fl-\Fm-\Fn}\,,\label{eq:m6}
\eea
where the mean $\mn{...}$ is taken within a single event, and the indices must be all different $i\neq j\neq k\neq\ell\neq\mu\neq\nu$. 
%
The two-particle azimuthal moment is composed of flow and nonflow ($\delta$)~\cite{Borghini_multipart,Borghini_multipart2}:
\be\M{2}=v^2+\delta\,.\label{eq:m2b}\ee
The event-average two-particle azimuthal moment is the two-particle flow anisotropy measurement:
\be\vv{2}^2\equiv\mM{2}=\mn{v^2}+\mn{\delta}\,.\label{eq:v2}\ee

To obtain the higher cumulants below, we assume flow and nonflow are uncorrelated and nonflow fluctuation is negligible as assumed in literature~\cite{v2method,Borghini_multipart,Borghini_multipart2}.
The four-particle azimuthal moment is~\cite{Borghini_multipart}
\be\M{4}=v^4+4v^2\delta+2\delta^2\,.\label{eq:m4b}\ee
The four-particle cumulant flow is given by~\cite{Borghini_multipart}
\be\vv{4}^4\equiv2\mM{2}^2-\mM{4}=2\mn{v^2}^2-\mn{v^4}\,.\label{eq:v4}\ee
%
The six-particle azimuthal moment is~\cite{Borghini_multipart,Borghini_multipart2,Bilandzic_directcalc}
\be\M{6}=v^6+9v^4\delta+18v^2\delta^2+6\delta^3\,.\label{eq:m6b}\ee
The six-particle cumulant flow is given by~\cite{Borghini_multipart,Borghini_multipart2,Bilandzic_directcalc}: 
\bea
\vv{6}^6&\equiv&\left(\mM{6}-9\mM{2}\mM{4}+12\mM{2}^3\right)/4\nonumber\\
&=&\left(\mn{v^6}-9\mn{v^2}\mn{v^4}+12\mn{v^2}^3\right)/4\,.\label{eq:v6}
\eea
%
We note that the left sides of Eqs.~(\ref{eq:v4}) and (\ref{eq:v6}) are notations by definition because the right sides can be negative. When the right sides are positive, then $\vv{4}$ and $\vv{6}$ are the four- and six-particle flow.

\section{Gaussian fluctuations assumption}

We now assume that the flow fluctuations are Gaussian,
\be\frac{dN}{dv}=\frac{1}{\sqrt{2\pi}\sigma}\exp\left[-\frac{(v-\mn{v})^2}{2\sigma^2}\right]\,.\label{eq:Gaus}\ee
Under the Gaussian fluctuation ansatz, we have
\bea
\mn{v^2}&=&\mn{v}^2+\sigma^2\,,\label{eq:g2}\\
\mn{v^4}&=&\mn{v}^4+6\mn{v}^2\sigma^2+3\sigma^4\,,\label{eq:g4}\\
\mn{v^6}&=&\mn{v}^6+15\mn{v}^4\sigma^2+45\mn{v}^2\sigma^4+15\sigma^6\,.\label{eq:g6}
\eea
Substituting them into Eqs.~(\ref{eq:v2}), (\ref{eq:v4}), and (\ref{eq:v6}), we obtain
\bw\bea
\vv{2}^2\equiv\mM{2}&=&\mn{v}^2+\sigma^2+\delta\,,\label{eq:v2g}\\
\vv{4}^4\equiv2\mM{2}^2-\mM{4}&=&(\mn{v}^2-\sigma^2)^2-2\sigma^4\,,\label{eq:v4g}\\
\vv{6}^6\equiv\left(\mM{6}-9\mM{2}\mM{4}+12\mM{2}^3\right)/4&=&\mn{v}^4(\mn{v}^2-3\sigma^2)\,.\label{eq:v6g}
\eea\ew
Eq.~(\ref{eq:v4g}) is as same as the result in Ref.~\cite{Voloshin_fluc,Ollitrault_fluc_nonflow} if the flow fluctuations are small so that $\sigma^4$ can be neglected. Under such circumstances, the flow fluctuation effect in $\vv{2}^2$ is $+\sigma^2$ (positive) and in $\vv{4}^2$ is $-\sigma^2$ (negative), as seen from Eqs.~(\ref{eq:v2g}) and (\ref{eq:v4g}), respectively.

In general flow fluctuation effects differ in different order azimuthal moments. Thus high order moments, coming with the unknown fluctuation effects, do not generally help resolving flow, flow fluctuation, and nonflow. When flow fluctuations are small, $\sigma^2/\mn{v}^2\ll1$, even under the Gaussian fluctuation ansatz, higher order moments do not add new information. This is because Eqs.~(\ref{eq:v4g}) and (\ref{eq:v6g}) would give the same information:
\be\vv{4}/\mn{v}\approx\vv{6}/\mn{v}\approx1-\sigma^2/2\mn{v}^2\,,\,\,{\rm if}\,\sigma^2/\mn{v}^2\ll1\,.\ee
Only up to the order of $\sigma^6/\mn{v}^6$ do \vv{6} and \vv{4} differ:
\be\vv{6}/\vv{4}\approx 1-\sigma^6/3\mn{v}^6\,,\,\,{\rm if}\,\sigma^2/\mn{v}^2\ll1\,.\label{eq:v6v4approx}\ee
Indeed data indicate that the six-particle $\vtwo{6}$ approximately equals to the four-particle $\vtwo{4}$~\cite{Tang,STAR_v2}. 

\begin{figure}[hbt]
\centerline{\includegraphics[width=0.4\textwidth]{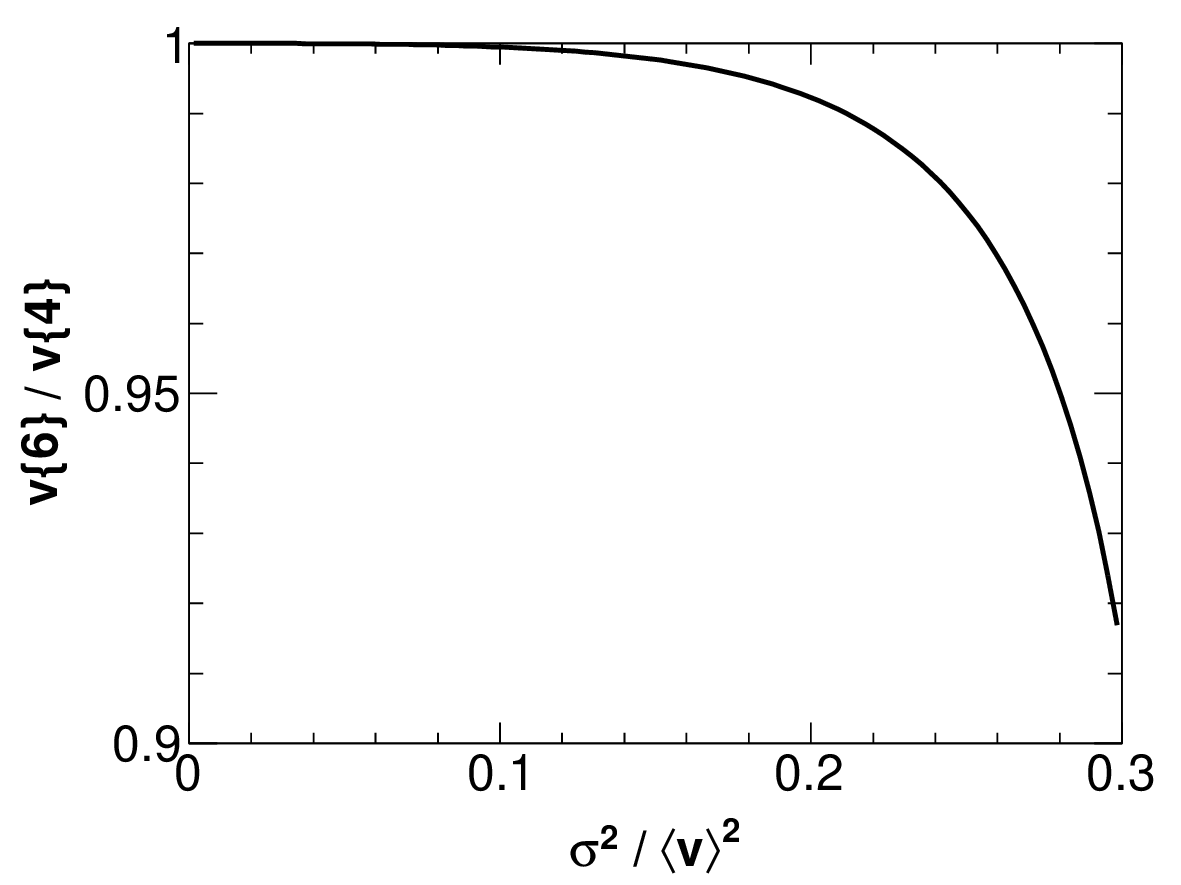}}
\caption{Ratio of $\vv{6}/\vv{4}$ as a function of $\sigma^2/\mn{v}^2$ under the Gaussian flow fluctuation ansatz.}
\label{fig}
\end{figure}

On the other hand, if flow fluctuations are large, 
then Eq.~(\ref{eq:v6g}) gives extra information than Eq.~(\ref{eq:v4g}). 
Figure~\ref{fig} illustrates this point where the ratio of $\vv{6}/\vv{4}$ is plotted against $\sigma^2/\mn{v}^2$:
\be\vv{6}/\vv{4}=\frac{(1-3\sigma^2/\mn{v}^2)^{1/6}}{(1-2\sigma^2/\mn{v}^2-\sigma^4/\mn{v}^4)^{1/4}}\,,\label{eq:v6v4}\ee
As seen, when $\sigma^2/\mn{v}^2\gtrsim0.2$, a reasonable accuracy measurement of $\vv{6}/\vv{4}$ would be able to determine the flow fluctuations. 
Note $\sigma^2/\mn{v}^2$ is roughly the relative difference between $\vv{2}$ and $\vv{4}$.

Odd harmonic anisotropies ($v_1$ and $v_3$) are dominated by fluctuations. The fluctuations and the averages are likely on the same order. In fact, the average $v_1$ and $v_3$ with respect to the reaction plane are zero, and this may imply that their fluctuations roughly equal to their averages with respect to their own harmonic axes. It is, therefore, hopeful that the triangular (and directed) flow, flow fluctuation, and nonflow can be uniquely determined by the three equations of Eqs.~(\ref{eq:v2g}), (\ref{eq:v4g}), and (\ref{eq:v6g}). 
We note, however, that in the case of small flow magnitude our neglect of the high order nonflow contributions 
may not be justified.

For even harmonic anisotropies, such as the $v_2$, the fluctuations are likely much smaller than the averages. In this case, the six-particle moment does not add much extra information beyond what is already in the four-particle moment. Thus, one will not be able to solve for flow, flow fluctuation, and nonflow from only two equations given by the two- and four-particle moments. One can hope to overcome this by increasing the event statistics so that small deviation in $\vv{6}/\vv{4}$ from unity may be measurable. However, it is possible that even with infinite statistics, a measurement of $\vv{6}/\vv{4}$ may not be enough to determine flow, flow fluctuation, and nonflow. This is because we have neglected those nonflow terms that are order(s) of magnitude smaller than the leading flow terms, but they may not be smaller than the high order terms in flow fluctuations. This depends on the details of nonflow, such as the multiplicities of multi-particle clusters, their anisotropies, and etc.


It is worthwhile to note that the reaction plane, or the participant plane, is not pertinent to the two- and multi-particle correlation measurements of flow. In our Gaussian fluctuation ansatz, the only assumption is that the flow is a Gaussian distribution about some mean $\mn{v}$; with respect to which plane the mean $\mn{v}$ arises is not central to our method. 
The idea of Gaussian flow fluctuation is not new. The possibility of extracting flow fluctuations under the Gaussian assumption via comparison of higher cumulants was studied earlier by Voloshin~\cite{Voloshin_ICHEP}. With relatively small fluctuation of $v_2$, the extraction of flow fluctuation is difficult. In this work we point out that in the case of large fluctuations (such as in $v_3$), extracting flow fluctuation may be possible under the Gaussian fluctuation ansatz.

In our Gaussian fluctuation model, the event-by-event $v$ may dynamically fluctuate to negative values. This may be naturally expected, for example, from fluctuations in the hydrodynamical responses~\cite{Stephanov} such that the maximal particle emission is no longer in the direction of the initial configuration minor axis. Experimentally, however, $v$ is often measured by the event-plane method where the event plane is reconstructed from final state particle momenta. The measured $v$ is then positive definite because of the redefinition of the event plane by the maximal emission direction of final state particles when it is orthogonal to the configuration space minor axis. The $v$ in our study, on the other hand, is determined by the cumulant (correlation) method. No prior plane is required by the cumulant method because only the particle pair-wise azimuthal angle difference is used. Given an event with particle momenta information, the flow cumulant is uniquely determined. It is made up by mean flow $\mn{v}$, flow fluctuations, and nonflow. Thus the $\mn{v}$ extracted from our method is not the one experimentally measured by final state particles, but the one with respect to some symmetry plane under the assumption of Gaussian fluctuations. This symmetry plane can be the true participant plane but the identification of this plane is not important in our decomposition method.

If in reality flow is not a Gaussian distribution with respect to any particular plane, then obviously our method will not work. For example, it was shown in Ref.~\cite{Voloshin_fluc} that Gaussian distributions of the $x$ and $y$ eccentricities in the reaction plane yield a Bessel-Gaussian distribution of the eccentricity magnitude in the participant plane. If elliptic flow is exactly proportional to such a participant plane eccentricity then the harmonic moments of the order of six and higher would give identical information as in the fourth order harmonic moment~\cite{Voloshin_fluc}. However, given an identical initial condition, the system response during collision evolution must also fluctuate~\cite{Stephanov}. Such response fluctuations could smear the final flow distribution towards Gaussian. If, on the other hand, flow is indeed Gaussian with respect to a particular plane, then the mean flow extracted from our method should be the $\mn{v}$ with respect to that plane. For elliptic flow, it is likely that this particular plane is the reaction plane, and the fluctuating participant plane gives rise to majority of the flow fluctuation. It is recently shown by Qiu and Heinz~\cite{Heinz} that the assumption of Bessel-Gaussian fluctuations works well for collisions of impact parameter $b<10$~fm but breaks down in more peripheral collisions, while the hypothesis of Gaussian fluctuations works better than the Bessel-Gaussian assumption for peripheral collisions, but breaks down for central collisions of $b<5$~fm. 

Experimentally, one may turn the argument around. The cumulants $\vv{2}$, $\vv{4}$, and $\vv{6}$ are uniquely determined regardless of which plane flow might be defined. Different models of flow fluctuations and mean $\mn{v}$ (i.e.~with respect to which plane) would give different relationships among the cumulants. It may therefore be possible to determine the nature of flow and flow fluctuations from experimental measurements of the flow cumulants. For example, if one measured $\vv{6}<\vv{4}$, then it could mean that the Bessel-Gaussian fluctuation model is not completely correct. It does not necessarily mean that our Gaussian fluctuation assumption is correct, but it could provide constraints to flow fluctuation models. If, on the other hand, one measured $\vv{6}>\vv{4}$, then neither the Bessel-Gaussian nor the Gaussian fluctuation model would be correct. One would have to devise new fluctuation model to accommodate experimental data. The preliminary data from STAR~\cite{LiYi} seem to favor our Gaussian fluctuation model.

\section{Toy-Model Simulations}

We verify our results by a toy-model {\em Monte Carlo} simulation. Below are the ingredients of the simulation.
\begin{itemize}
\item For each event, we generate a $v_2$ value according to the Gaussian distribution of Eq.~(\ref{eq:Gaus}) with $\mn{v_2}$ and $\sigma_2$. We do the same for $v_3$ with $\mn{v_3}$ and $\sigma_3$. We allow negative $v_2$ and of course negative $v_3$. We require $|v_n|<0.5$.
\item We set the reaction plane angle to zero. We generate triangularity axis angle ($\psi_3$) randomly between 0 and $2\pi$. We generate $N-\ncl{2}$ particles with azimuthal angle $\phi=0$-$2\pi$ according to the flow modulation of Eq.~(\ref{eq:fourier}):
\be dN/d\phi\propto1+2v_2\cos2\phi+2v_3\cos3(\phi-\psi_3)\,.\ee
\item $\ncl{2}$ of those $N-\ncl{2}$ particles each has a partner with identical azimuthal angle (so total number of particles is $N$, and $\ncl{2}\leq N/2$). These correlated pairs yield a nonflow correlation, which is simply given by $\delta=2\ncl{2}/N(N-1)$. There are no clusters of more than two particles. The total particle multiplicity $N$ and the number of directly correlated pairs $\ncl{2}$ are fixed for all events; there are no multiplicity fluctuations or nonflow fluctuations in our simulation. 
\end{itemize}

We note that the analysis of $v_2$ and $v_3$ are independent of each other, so one may simulate events with only $v_2$ or $v_3$ (with corresponding fluctuations and nonflow). We have simulated several cases, with both finite $v_2$ and $v_3$, only finite $v_2$, or only finite $v_3$. We obtained the same results.


Table~\ref{tab} lists the inputs and our results from several simulations. The values of the multiplicity $N$ in Cases (i) and (ii) correspond to real data of roughly 30-40\% centrality of 200 GeV Au+Au collisions at RHIC~\cite{Levente}. Those in Cases (iii) and (iv) correspond to real data of roughly the top 5\% centrality of 200 GeV Au+Au collisions at RHIC~\cite{Levente}. The $v_2$ values in Case (i) and (iii) correspond approximately those measured experimentally in the respective centralities~\cite{STAR_v2}. We used a relatively small $v_2$ fluctuation for Case (i) and large $v_2$ fluctuation for Case (iii). Note $\sigma_2^2/\mn{v_2}^2\approx0.2$ implies an approximately 20\% difference between $\vv{2}$ and $\vv{4}$. We set the $\ncl{2}$ value such that the input nonflow and flow fluctuation are on the same order, $\delta\sim\sigma^2$. 

\begin{table*}
\caption{Toy-model simulation to verify our method. Several cases are studied:
(i) $N=400, \ncl{2}=72, \mn{v_2}=0.07, \sigma_2=0.03, \delta_2=9\times10^{-4};$  
(ii) $N=400, \ncl{2}=72, \mn{v_3}=0, \sigma_3=0.03, \delta_3=9\times10^{-4};$  
(iii) $N=1200, \ncl{2}=288, \mn{v_2}=0.03, \sigma_2=0.02, \delta_2=4\times10^{-4};$  
(iv) $N=1200, \ncl{2}=288, \mn{v_3}=0.03, \sigma_3=0.03, \delta_3=4\times10^{-4}.$  
We simulated $10^7$ events for each case.
}
\label{tab}
\begin{ruledtabular}
\begin{tabular}{cccccc}
Case
& (i) 
& (ii) 
& (iii)
& (iv) 
\\
$\mM{2}$ 
& $(6.698\pm0.003)\times10^{-3}$ 
& $(1.803\pm0.001)\times10^{-3}$ 
& $(1.7002\pm0.0008)\times10^{-3}$ 
& $(2.199\pm0.001)\times10^{-3}$ 
\\
$\mM{4}$ 
& $(7.514\pm0.007)\times10^{-5}$ 
& $(7.29\pm0.02)\times10^{-6}$ 
& $(5.849\pm0.007)\times10^{-6}$ 
& $(1.127\pm0.001)\times10^{-5}$ 
\\
$\mM{6}$ 
& $(1.139\pm0.002)\times10^{-6}$ 
& $(4.79\pm0.03)\times10^{-8}$ 
& $(2.955\pm0.008)\times10^{-8}$ 
& $(8.95\pm0.02)\times10^{-8}$ 
\\
$\vv{2}$ 
& $0.08184\pm0.00002$ 
& $(0.04246\pm0.00002)$ 
& $(0.04123\pm0.00001)$ 
& $(0.04689\pm0.00001)$ 
\\
$\vv{4}^4$ 
& $(0.0618\pm0.0001)^4$ 
& $(-7.9\pm0.2)\times10^{-7}$ 
& $(-6.8\pm0.9)\times10^{-8}$ 
& $(-1.59\pm0.02)\times10^{-6}$ 
\\
$\vv{6}^6$ 
& $(0.0615\pm0.0003)^6$ 
& $(0.0\pm0.1)\times10^{-9}$ 
& $(-2.4\pm0.3)\times10^{-10}$ 
& $(-1.46\pm0.08)\times10^{-9}$ 
\\
$\mn{v}$ 
& $0.070\pm0.003$ 
& $0.007\pm0.009$ 
& $0.0300\pm0.0009$ 
& $0.0300\pm0.0006$ 
\\
$\sigma$ 
& $0.030\pm0.004$ 
& $0.029\pm0.002$ 
& $0.0200\pm0.0005$ 
& $0.0300\pm0.0001$ 
\\
$\delta$ 
& $(8.4\pm6.1)\times10^{-4}$ 
& $(9.1\pm0.2)\times10^{-4}$ 
& $(4.0\pm0.8)\times10^{-4}$ 
& $(4.0\pm0.4)\times10^{-4}$ 
\end{tabular}
\end{ruledtabular}
\end{table*}

We use the Q-vector method~\cite{Bilandzic_directcalc} 
to calculate the two-, four-, and six-particle azimuthal moments from the simulated events. The calculated moments are listed in Table~\ref{tab}. Also listed are the calculated $\vv{2}$, $\vv{4}$, and $\vv{6}$ values for comparison. 

Using Eqs.~(\ref{eq:v2g}), (\ref{eq:v4g}), and (\ref{eq:v6g}), we can solve for flow, flow fluctuation, and nonflow. We can either do it numerically, or use the $\chi^2$ minimization method. In order to obtain the proper errors on the solutions, we use the latter. We treat the statistical errors on the moments as independent. We compute the normalized $\chi^2$, the quadratic sum of the differences between the left and right sides of Eqs.~(\ref{eq:v2g}), (\ref{eq:v4g}), and (\ref{eq:v6g}). 
We minimize the $\chi^2$ to obtain the solutions of $\mn{v}$, $\sigma$, and $\delta$; the minimum $\chi^2$ should in principle be zero as is the case in our solutions. 
The solutions are tabulated in Table~\ref{tab}. As seen from the table, the resolved flow, flow fluctuation, and nonflow compare well to the inputs.
We found, when the flow fluctuations are small, $\sigma^2/\mn{v}^2\ll1$, that the fitted $\mn{v}$ and $\sigma$ are strongly correlated with each other, and the fitted $\delta$ is strongly anti-correlated. On the other hand, when the flow fluctuations are large, the covariances of the error matrix nearly vanish and the fitted $\mn{v}$, $\sigma$, and $\delta$ are mostly independent.



We simulated $10^7$ events for each case. As seen, the errors on $\sigma$ and $\delta$ are the largest for Case (i). This is because of the smaller $\sigma^2/v^2$ in this case. As mentioned earlier, small $\sigma^2/v^2$ makes the four- and six-particle moments not much different. This results in a correlated errors on $v^2$ and $\sigma^2$ that are not well controlled and may become large. This in turn gives a large error on $\delta$.

We used a simple nonflow model of particle pairs of identical azimuthal angle (i.e.~a $\delta$-function in two-particle correlations). Our decomposition method does not depend on the details of the nonflow model, but only the magnitude of the nonflow contribution to two-particle correlations. We have tested our method using other two-particle correlation shapes for nonflow and found our method is robust.

\section{Differential Flow}

In this section we discuss how to obtain differential flow, e.g.~as a function of $p_T$ or particle species. For easy discussion we will use $p_T$ as an example, but our discussion is general for any type of differential flow. 

In order to obtain differential flow, one can form azimuthal moments with one particle within the interested $p_T$ region and the other particle(s) over the unrestricted $p_T$ region. These azimuthal moments are identical to those in Eqs.~(\ref{eq:m2}), (\ref{eq:m4}), and (\ref{eq:m6}), except that the first azimuthal angle $\phi_i$ is replaced by that of the particle within the interested $p_T$ region, $\Fp$.
\bea
\M{2'}&=&\mne{\Fp-\Fj}\,,\label{eq:m2p}\\
\M{4'}&=&\mne{\Fp+\Fj-\Fk-\Fl}\,,\label{eq:m4p}\\
\M{6'}&=&\mne{\Fp+\Fj+\Fk-\Fl-\Fm-\Fn}\,.\label{eq:m6p}
\eea
Similar to Eqs.~(\ref{eq:m2b}), (\ref{eq:m4b}) and (\ref{eq:m6b}), we have
\bea
\M{2'}&=&v'v+\delta'\,,\label{eq:m2pb}\\
\M{4'}&\approx&v'v(v^2+2\delta)+2\delta'(v^2+\delta)\,,\label{eq:m4pb}\\
\M{6'}&\approx&v'v(v^4+6v^2\delta+6\delta^2)+\nonumber\\
&&3\delta'(v^4+4v^2\delta+2\delta^2)\,.\label{eq:m6pb}
\eea

Suppose the anisotropic flow within the interested $p_T$ region is $v'$ with Gaussian fluctuation $\sigma'$. We will assume that $v'$ is completely correlated with $v$. That is, $v'$ and $v$ in the same event must satisfy:
\be(v'-\mn{v'})/\sigma'=(v-\mn{v})/\sigma\,.\ee
The probability of an event having $v'$ and $v$ is given by Eq.~(\ref{eq:Gaus}).
Analogous to Eqs.~(\ref{eq:g2}), (\ref{eq:g4}), and (\ref{eq:g6}), we have
\bea
\mn{v'v}&=&\mn{v'}\mn{v}+\sigma'\sigma\,,\\
\mn{v'v^3}&=&\mn{v'}\mn{v}\left(\mn{v}^2+3\sigma^2\right)+\nonumber\\
&&3\sigma'\sigma\left(\mn{v}^2+\sigma^2\right)\,,\\
\mn{v'v^5}&=&\mn{v'}\mn{v}\left(\mn{v}^4+10\mn{v}^2\sigma^2+15\sigma^4\right)+\nonumber\\
&&5\sigma'\sigma\left(\mn{v}^4+6\mn{v}^2\sigma^2+3\sigma^4\right)\,.
\eea

Similar to Eqs.~(\ref{eq:v2g}), (\ref{eq:v4g}) and (\ref{eq:v6g}), 
we have
\bw\bea
\mM{2'}
&=&\mn{v'v}+\delta'\nonumber\\
&=&\mn{v'}\mn{v}+\sigma'\sigma+\delta'\,,\label{eq:v2pg}\\
2\mM{2'}\mM{2}-\mM{4'}
&\approx&2\mn{v'v}\mn{v^2}-\mn{v'v^3}\nonumber\\
&=&\mn{v'}\mn{v}(\mn{v}^2-\sigma^2)-\sigma'\sigma(\mn{v}^2+\sigma^2)\,,\label{eq:v4pg}\\
\left(\mM{6'}-3\mM{2'}\mM{4}-6\mM{2}\mM{4'}+12\mM{2'}\mM{2}^2\right)/4
&\approx&\left(\mn{v'v^5}-3\mn{v'v}\mn{v^4}-6\mn{v^2}\mn{v'v^3}+12\mn{v'v}\mn{v^2}^2\right)/4\nonumber\\
&=&\mn{v}^3(\mn{v'}\mn{v}^2-2\mn{v'}\sigma^2-\mn{v}\sigma'\sigma)\,.\label{eq:v6pg}
\eea\ew
Here we have taken the nonflow correlation effect between one particle in the interested $p_T$ region and another particle in the unrestricted $p_T$ region to be $\delta'$. 

Knowing the $\mn{v}$, $\sigma$, and $\delta$ for particles in the unrestricted $p_T$ region, we can readily solve for $\mn{v'}$, $\sigma'$ and $\delta'$ from Eqs.~(\ref{eq:v2pg}), (\ref{eq:v4pg}), and (\ref{eq:v6pg}).

\section{Conclusions}

Anisotropic flow measurements have provided valuable information about the early stage of relativistic heavy-ion collisions. The contamination of nonflow, however, has hampered precise extraction of the hydrodynamic properties of the collision system. Nonflow and flow fluctuations have not been successfully separated in experimental measurements. We argue, if the flow fluctuations are Gaussian, that it may be possible to disentangle flow, flow fluctuation, and nonflow by simultaneous measurements of two-, four-, and six-particle azimuthal moments. We demonstrate that this possibility is real when 
the flow fluctuations are sizable relative to the average flow magnitude. The disentanglement may be difficult when 
the relative flow fluctuations are small. We have verified our conclusions by a toy-model simulation. 

To conclude, our study suggests that it may be possible to measure elliptic and triangular flows, flow fluctuations, and nonflow uniquely in experiment. If, on the other hand, real data analysis does not yield reliable results, our study offers additional insight--It may mean that either the Gaussian fluctuation assumption is not fully applicable and/or the flow fluctuations are too small.

\section*{Acknowledgment}

We thank Ante Bilandzic and Raimond Snellings for useful communications. This work is supported by U.S. Department of Energy under Grants DE-AC02-98CH10886, DE-FG02-88ER40412, and DE-FG02-89ER40531. AT and FW thank the Institute of Particle Physics, Central China Normal University, Wuhan where the 3$^{\rm rd}$ Asia Triangle Heavy-Ion Conference was held during which the initial general idea of the work was materialized.

\end{document}